\documentclass[twocolumn,amsmath,pra,nobibnotes,showpacs,superscriptaddress]{revtex4}
\usepackage{graphicx}

\begin{document}
\title{Remote preparation of arbitrary time-encoded single-photon ebits}

\author{Alessandro Zavatta}
\affiliation{Istituto Nazionale di Ottica Applicata - CNR, L.go E. Fermi, 6, I-50125,
Florence, Italy}

\author{Milena D'Angelo}
\affiliation{LENS, Via Nello Carrara 1, 50019 Sesto Fiorentino, Florence, Italy}

\author{Valentina Parigi}
\affiliation {Department of Physics, University of Florence, I-50019 Sesto Fiorentino,
Florence, Italy}

\author{Marco Bellini}
\email{bellini@inoa.it} \affiliation{Istituto Nazionale di Ottica Applicata - CNR, L.go
E. Fermi, 6, I-50125, Florence, Italy} \affiliation{LENS, Via Nello Carrara 1, 50019
Sesto Fiorentino, Florence, Italy}

\date{\today}

\begin{abstract}
We propose and experimentally verify a novel method for the remote preparation of
entangled bits (ebits) made of a single-photon coherently delocalized in two
well-separated temporal modes. The proposed scheme represents a remotely tunable source
for tailoring arbitrary ebits, whether maximally or non-maximally entangled, which is
highly desirable for applications in quantum information technology. The remotely
prepared ebit is studied by performing homodyne tomography with an ultra-fast balanced
homodyne detection scheme recently developed in our laboratory.
\end{abstract}


\maketitle

Entanglement, nonlocal correlations, indistinguishable alternatives are, historically,
among the most intriguing and appealing topics of quantum mechanics. Besides their
relevance in fundamental physics \cite{epr}, these phenomena have attracted much
attention due to their usefulness in quantum information technology \cite{qu_infor}.
Extravagant but promising protocols such as quantum teleportation, quantum
cryptography, and quantum computation have been proposed and experimentally verified
(see, e.g., \cite{qu_infor} and references therein). All these schemes were originally
based on two-photon entanglement. Recently, increasing attention has been given to a
new quantum information perspective: the carriers of quantum information are no longer
the photons, but rather the field modes ``carrying'' them. Based on this idea, two
different approaches have been followed. The first one exploits the entanglement in
momentum generated when a single-photon impinges on a beam-splitter and is
characterized by the state $\alpha |1\rangle_a |0\rangle_b + \beta |0\rangle_a
|1\rangle_b$, where $a$ and $b$ denote two distinct spatial modes, $\alpha$ and $\beta$
are complex amplitudes such that $|\alpha|^2+|\beta|^2=1$ (see, e.g.,
\cite{ent_1phot,knill} and references therein). The second and more recent road has
been traced by Gisin's group \cite{gisin} on the line of Franson's approach
\cite{franson}, and leads to two-photon systems entangled in ultra-short co-propagating
temporal modes (or ``time-bins'') \cite{simon}.

In this Letter, we propose the first remotely tunable source of arbitrary single-photon
entangled states (ebits) in the time domain and experimentally demonstrate its working
principle. We start from the spontaneous parametric down conversion (SPDC) signal-idler
pairs~\cite{klyshko} generated by a train of phase-locked pump
pulses~\cite{spdc_pulse,gisin} and generate indistinguishability between pairs of
consecutive non-overlapping temporal modes propagating in the idler channel; this
enables us to remotely delocalize the twin signal photon between two identical and
well-separated time-bins, thus generating the single-photon temporal ebit: $\alpha
|1^{(n)}\rangle |0^{(n+1)} \rangle + \beta |0^{(n)} \rangle |1^{(n+1)} \rangle$, where
$n$ denotes the temporal mode associated with the $n^{th}$ pump pulse. Both maximally
and non-maximally entangled single-photon states, with any relative phase, can be
produced by performing simple and reversible operations in the remote idler channel.
The proposed scheme may find immediate application in quantum information technology;
single-photon ebits have been proven to enable linear optics quantum teleportation
\cite{ent_1phot,gisin_tele} and play a central role in linear optics quantum
computation \cite{knill,pittman}. Furthermore, time-bin entanglement has been proven
suitable for long distance applications \cite{gisin_crypto,gisin_tele}, where the
insensitivity to both depolarization and polarization fluctuations becomes a strong
requirement. In addition, since the carriers of entanglement are naturally separated
(i.e., no further optical element is required) and undergo the same losses,
entanglement in time is less sensitive to losses and easier to purify \cite{yamamoto}.

\begin{figure}[t]
\includegraphics [width=2.6in]{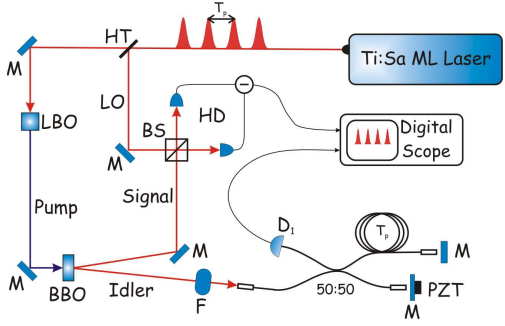}
\caption[] {\label{fig_setup}(Color online) Schematic representation of the
experimental setup. 50:50 is a 3 dB fiber coupler, HT a high-transmission beam
splitter, and M are mirrors. See text for further details.}
\end{figure}

The experimental setup is pictured in Fig.~\ref{fig_setup}. The $1.5$ ps pulses at
$786$ nm from a mode-locked Ti:Sapphire laser at a repetition rate of $82$ MHz are
frequency doubled in a LBO crystal. The resulting pulse train impinges on a non-linear
BBO crystal cut for degenerate ($\Omega_s=\Omega_i=\Omega_p/2$) non-collinear type-I
SPDC; signal-idler photon pairs centered around $786$ nm are thus generated in two
distinct spatial modes. A single mode fiber and a pair of etalon interference filters
(F) are employed for spatial and spectral filtering of the idler beam before its
entrance in a fiber-coupled piezo-controlled (PZT) Michelson interferometer; a
single-photon detector ($D_1$) is inserted at the exit port of the interferometer. The
signal beam propagates in free space before being mixed at a 50-50 beam-splitter (BS)
with a local oscillator (LO) for high-frequency time-domain balanced homodyne detection
(HD) \cite{josa02,marco_pra70_04}.
Spatial and spectral filtering of the idler mode guarantees the conditional projection
of the signal photons into a single-photon pure state
\cite{Aichele_02,marco_prl90_03,marco_pra69_04}. On the other hand, the Michelson
interferometer generates indistinguishability between two consecutive temporal modes
propagating in the idler channel: an idler photon detected by $D_1$ may have been
generated by either the $N^{th}$ or the $(N+1)^{th}$ pump pulse, provided that the time
delay ($T$) between the short and long arms of the interferometer is chosen to be
approximately equal to the time separation between two consecutive pump pulses
($T_p=12.3$ ns). Notice that the bandpass of the spectral filter in the idler arm
($\sigma_i=50$ GHz) is wide enough so that no first order interference occurs
($\sigma_i \gg \pi / T_p$).

Based on a standard quantum mechanical calculation, we find that the combination of
indistinguishability and tight filtering in the idler channel allows the conditional
remote preparation, in the signal channel, of the temporally delocalized single-photon
ebit:
\begin{equation}\label{state_fin}
|\Psi_s^{\phi_i} \rangle = \frac{1}{\sqrt{2}} (|1^{(n)}, 0^{(n+1)} \rangle + e^{-i
\phi_i} |0^{(n)}, 1^{(n+1)} \rangle),
\end{equation}
with $\phi_i= \Omega_p(T_p-T/2)$. Interestingly, the relative phase $\phi_i$
characterizing the remotely prepared ebit is defined not only by the phase difference
introduced by the Michelson interferometer ($\varphi_{int}=\Omega_i T$), but also by
the relative phase between consecutive pump pulses ($\varphi_{pump}=\Omega_p T_p$). The
result of Eq.~(\ref{state_fin}) represents the temporal counterpart of the spatially
delocalized single-photon produced at the output ports of a beam splitter; this case
has been studied experimentally by Babichev, {\em et al.} \cite{lvovsky_prl04_2}.
However, different from Ref.~\cite{lvovsky_prl04_2}, the entangled state of
Eq.~(\ref{state_fin}) has been prepared remotely, without performing any manipulation
on the signal photons. It is the interferometer in the idler arm which generates
indistinguishability between two consecutive non-overlapping temporal modes; this
indistinguishability, together with the coherence of both pump beam and SPDC process,
gives rise, in the signal channel, to the coherent superposition of two previously
independent and still temporally separated time-bins. An important advantage of such a
remote state preparation scheme is the possibility of generating both maximally and
non-maximally single-photon entangled states, with any relative phase $\phi_i$, by
performing simple and reversible operations in the idler arm (or on the train of pump
pulses). For instance, two of the four Bell states, namely $|\Psi_s^{\pm} \rangle =
\frac{1}{\sqrt{2}} (|1^{(n)}, 0^{(n+1)} \rangle \pm |0^{(n)}, 1^{(n+1)} \rangle)$, can
be easily generated by manipulating the interferometer. Furthermore, the probability
amplitudes characterizing the delocalized single-photon may be continuously varied by
simply introducing controllable losses in one arm of the interferometer; this has the
only effect of lowering the production rate but does not introduce any impurity in the
state generated in the signal channel.

The expected two-mode Wigner function \cite{leonardth} for the delocalized
single-photon of Eq.~(\ref{state_fin}) is given by:
\begin{eqnarray}\label{wigner_fin}
& & W^{\phi_i}(x_1,y_1; x_2,y_2)= \frac{1}{2}[8 W^{\phi_i}_{10} (x_1,y_1; x_2,y_2) \;
\\ &+&  W_1(x_1,y_1) W_0(x_2,y_2) + W_0(x_1,y_1) W_1(x_2,y_2)],\nonumber
\end{eqnarray}
where $W_1(x,y)=\frac{2}{\pi} e^{-2x^2} e^{-2y^2} (4x^2+4y^2-1)$ and
$W_0(x,y)=\frac{2}{\pi} e^{-2x^2} e^{-2y^2}$ are the single-mode Wigner functions
associated with a single-photon Fock state and with the vacuum, respectively; on the
other hand, $W^{\phi_i}_{10} (x_1,y_1; x_2,y_2)$ is a non-factorable 4-D function which
couples the quadratures of two consecutive non-overlapping signal temporal modes:
\begin{eqnarray}\label{wigner_acc}
W^{\phi_i}_{10} (x_1,y_1; x_2,y_2)&=& W_0(x_1,y_1) W_0(x_2,y_2) \nonumber \\ & \times&
(x_1 x_2^{\phi_i}+y_1 y_2^{\phi_i}),
\end{eqnarray}
where $x_2^{\phi_i}=x_2 \cos \phi_i -y_2 \sin \phi_i$ and $y_2^{\phi_i}=x_2 \sin \phi_i
+y_2 \cos \phi_i$. Then, the Wigner function associated with the delocalized signal
photon contains information about the characteristic phase $\phi_i$ introduced through
the idler arm. Also notice that, by introducing the phase-dependent correlation
quadratures $x_{\pm}^{\phi_i} = (x_1 \pm x_2^{\phi_i})/\sqrt{2}$, and $y_{\pm}^{\phi_i}
= (y_1 \pm y_2^{\phi_i})/\sqrt{2}$, the Wigner function of Eq.~(\ref{wigner_fin})
factors:
\begin{eqnarray}\label{wigner_finAB}
W(x_+^{\phi_i} ,y_+^{\phi_i} ; x_-^{\phi_i} ,y_-^{\phi_i} )= W_1(x_+^{\phi_i}
,y_+^{\phi_i} ) \times W_0(x_-^{\phi_i} ,y_-^{\phi_i} ).
\end{eqnarray}
This result explicitly indicates that the delocalized single-photon cannot be described
in terms of the quadratures associated with neither one of the two distant temporal
modes ($1$ and $2$), separately; however, the single-photon is well defined in the
phase space $(x_+^{\phi_i},y_+^{\phi_i})$, while the vacuum is defined in the phase
space $(x_-^{\phi_i},y_-^{\phi_i})$. Thus, the 4-D Wigner function reproduces the
correlations remotely generated between pairs of well-separated temporal modes in the
signal arm.

We have experimentally verified the correctness of the above predictions by performing
balanced homodyne tomography and reconstructing both the density matrix and the Wigner
function of the ebit remotely prepared in the signal channel. The density matrix has
been reconstructed directly from the homodyne data by employing the method developed by
D'Ariano, {\em et al.} \cite{dariano_den}; its elements have then been used to
reconstruct the Wigner function (for more details see our previous works
\cite{marco_pra70_04,marco_science_04}).

In order to reconstruct the two-mode 4-D Wigner function of Eq.~(\ref{wigner_fin}), one
would normally need to measure the joint marginal distribution of the quadratures
$X_1(\theta_1)=x_1 \cos \theta_1 - y_1 \sin \theta_1$ and $X_2(\theta_2)=x_2 \cos
\theta_2 - y_2 \sin\theta_2$, while varying the phases $\theta_1$ and $\theta_2$ of two
LO pulses spatially and temporally matched (i.e., synchronized) to the modes $1$ and
$2$, respectively. However, the particular state investigated here is invariant with
respect to the global phase, and only the relative phase $\Delta \theta
=\theta_1-\theta_2$ needs to be controlled in the experiment
\cite{lvovsky_prl04_2,leonardth}. Moreover, the joint marginal distribution is
invariant under interchange of $\phi_i$ and $\Delta \theta$. We exploited this property
in order to overcome the difficulty connected to the generation of a pair of
phase-controllable LO pulses out of the train coming from the laser. Rather than
varying the relative LO phase, one may keep $\Delta \theta$ fixed (by just using any
two consecutive pulses directly from the mode-locked train) and vary the phase $\phi_i$
by means of the interferometer. Although what we actually do in this case is to measure
fixed quadratures on the two modes for a varying quantum state $|\Psi_s^{\phi_i}
\rangle$, it is immediate to show that this is equivalent to performing a conventional
LO phase scan of the fixed quantum state $|\Psi_s^{\phi_i=const} \rangle$. We shall
name this technique as ``remote balanced homodyne tomography''.

For each value of the interferometer phase $\varphi_{int}$ and upon detection of an
idler photon, stable and fast quadrature measurements have been realized on the
corresponding pair of consecutive signal time-bins (plus one containing just the vacuum
and used for calibration), while keeping both the local oscillator and the homodyne
detection apparatus unchanged. A total of $10^6$ quadrature measurements, equally
distributed over the range $[0,\pi]$ of $\varphi_{int}$, has been performed on each of
the three time-bins.
\begin{figure}[t]
\includegraphics [width=3.4 in]{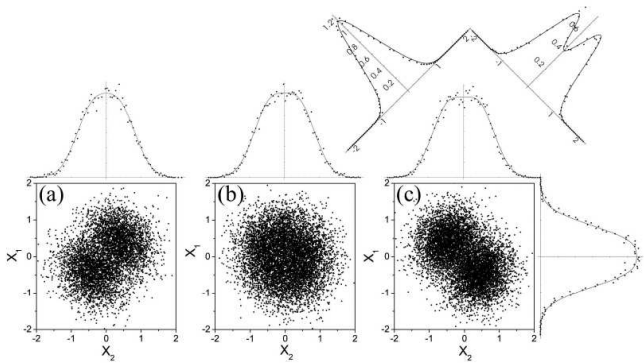}
\caption[] {\label{fig_correlation} Joint marginal distributions of the measured
two-mode field quadratures for: (a) $\phi_{i}=-\Delta \theta$, (b)
$\phi_{i}=\pi/2-\Delta \theta$, (c) $\phi_{i}=\pi-\Delta \theta$, while leaving $\Delta
\theta$ fixed. These are also the joint marginal distributions $p(X_1,X_2,\Delta
\theta)$ associated with the ebit of Eq.~(\ref{state_fin}) for $\phi_i=0$, and
corresponding, respectively, to $\Delta \theta = 0$, $\pi/2$, $\pi$. The histograms are
the single-mode marginal distributions $p(X_1)$ and $p(X_2)$ together with the
corresponding best fits. The marginals for the $x_{\pm}$ quadratures are plotted on the
diagonal axes above (c).}
\end{figure}
The experimental results are reported in Fig.~\ref{fig_correlation}, where we plot the
measured values of the quadratures $X_1$ and $X_2$ obtained for three different values
of the remote phase $\phi_i$, while leaving $\Delta \theta$ fixed. According to the
above reasoning, these results also represent the marginal distributions
$p(X_1,X_2,\Delta \theta)$ associated with the ebit of Eq.~(\ref{state_fin}) for
$\phi_i=0$, and obtained for three different values of the relative phase $\Delta
\theta$. Notice that, while the joint distribution $p(X_1,X_2,\Delta \theta)$ is
strongly phase-dependent, the marginal distributions $p(X_1)$ and $p(X_2)$ associated
with each temporal mode, separately, are phase-independent. This is consistent with the
fact that each mode, separately, is an incoherent statistical mixture of vacuum and
single-photon Fock state; however, the pair of modes $1$ and $2$, as a whole, is in the
coherent superposition described by Eq.~(\ref{state_fin}), with $\phi_i=0$. Figure
\ref{fig_correlation} also shows that a single-photon Fock state is defined in the
phase space $(x_+^{\phi_i=0},y_+^{\phi_i=0})$, while the vacuum is defined in the phase
space $(x_-^{\phi_i=0 },y_-^{\phi_i=0})$, as expected from Eq.~(\ref{wigner_finAB}).

\begin{figure}[t]
\includegraphics [width=2.6in]{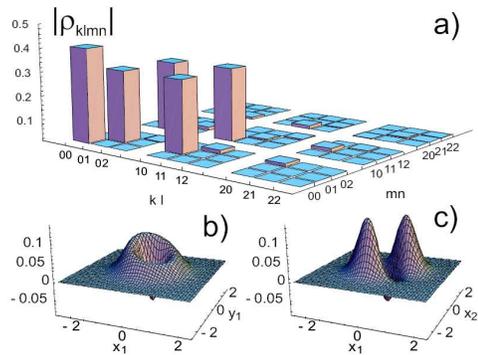}
\caption[] {\label{fig_wigner}(Color online) (a) Reconstructed density matrix  elements
$\rho_{klmn}=\langle k_1 l_2 | \hat \rho | m_1 n_2\rangle$ corresponding to the state
of Eq.~(\ref{state_fin}) with $\phi_i=0$. Cross sections of the reconstructed 4-D
Wigner function: (b) $W^{\phi_i=0}(x_1,y_1; -0.1 ,-0.1)$, and (c) $W^{\phi_i=0}(x_1,0;
x_2,0)$. }
\end{figure}

Figure~\ref{fig_wigner} (a) reports the reconstructed density matrix:  $\hat{\rho}=
(1-\eta)|0\rangle \langle0|+\eta|\Psi_s^{\phi_i=0}\rangle \langle \Psi_s^{\phi_i=0}|$,
where the overall efficiency $\eta=60.5$\% accounts for both preparation and detection
efficiencies; notice that almost no multi-photon contribution exists. From this figure
it is also apparent that the vacuum contamination, hence the losses, does not degrade
the coherence of the remotely delocalized single-photon; in fact, both the non-diagonal
and the diagonal ($|01 \rangle \langle 01|$ and $|10 \rangle \langle 10|$) elements of
the reconstructed density matrix are reduced by the same amount. This may be understood
as a consequence of the common losses undergone by the pair of entangled time-bins.
Figures \ref{fig_wigner}(b) and (c) reproduce, respectively, the $(x_1,y_1)$ and
$(x_1,x_2)$ sections of the reconstructed 4-D Wigner function $W^{\phi_i=0}(x_1, y_1,
x_2, y_2)$. The cross section $(x_1,y_1)$ resembles the standard Wigner function of a
single-photon Fock state, but is characterized by a well-defined phase; the existence
of this phase is the result of the coherent delocalization of the single-photon between
two separate temporal modes. The $(x_1,x_2)$ section of the reconstructed Wigner
function explicitly shows the correlation between the quadratures $x_1$ and $x_2$; the
non-factorable nature of the delocalized single-photon is here apparent.

In summary, the experimental reconstruction of the Wigner function of the conditionally
prepared single-photon ebit has enabled us to verify its entangled nature and study its
purity. Beside the non-classical behavior typical of single-photon Fock states
(negative values around the origin), the reconstructed 4-D Wigner function has been
found to be characterized by an intriguing phase information and by correlation between
well separated temporal modes, as expected from Eq.~(\ref{wigner_fin}). It may seem
counterintuitive that a single photon simultaneously affects two non-overlapping
temporal modes or, equivalently, carries a well defined phase. However, the effect is a
direct consequence of the coherent superposition remotely induced between otherwise
independent signal time-bins; it can then be understood in terms of quantum
entanglement between two co-propagating but distinct temporal modes carrying a
single-photon.

\begin{figure}[]
\includegraphics [width=3 in]{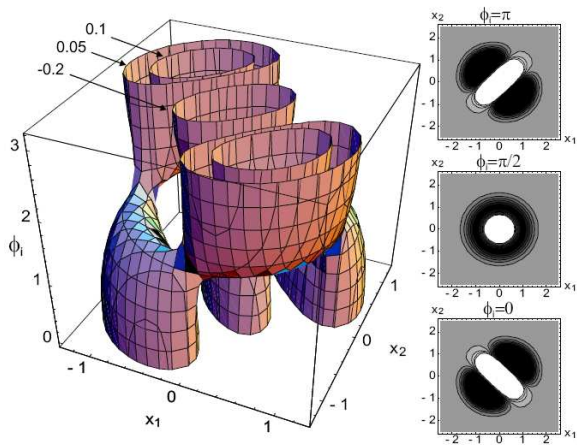}
\caption[] {\label{fig_contour} (Color online) 3-D contour plot of the Wigner function
section $W^{\phi_i}(x_1,0,x_2,0)$ associated with the single-photon ebit of
Eq.~(\ref{state_fin}) versus its characteristic remotely tunable phase $\phi_i$. The
surfaces shown correspond to three values of the Wigner function, namely: $W^{\phi_i}=
\; -0.2, \; 0.05, \; 0.1$. Insets: contour plots for three specific values of the phase
$\phi_i$.}
\end{figure}

From the applicative viewpoint, one of the most interesting aspects of the proposed
scheme is the dependence of the relative phase characterizing the delocalized (signal)
single photon on both the relative phase between pump pulses and the phase delay
introduced by the remote Michelson interferometer. Based on this effect, for any fixed
value of the remotely controlled phase $\phi_i$, one may generate, in the signal arm, a
specific single-photon ebit. This point is pictorially demonstrated by
Fig.~\ref{fig_contour}, where we draw the contour plots of the $(x_1,x_2)$ section of
the 4-D Wigner function for all possible values of the remotely tunable phase $\phi_i$
characterizing the ebit of Eq.~(\ref{state_fin}). The Wigner function
$W^{\phi_i}(x_1,0;x_2,0)$ associated with each conditionally prepared ebit
$|\Psi_s^{\phi_i}\rangle$ reveals a specific correlation between the field quadratures
of two distinct signal temporal modes; as the preparation phase $\phi_i$ is changed
from $0$ to $\pi$, we observe the transition from correlated to anti-correlated
quadratures through a ``saddle" point at $\phi_i = \pi/2$, where the anti-correlation
is transferred into the quadrature space $(x_1,y_2)$ (not shown in figure). In other
words, the proposed scheme can be regarded as a remotely tunable source of arbitrary
single-photon ebits; such a source is highly desirable for applications in quantum
information technology.

This work has been performed in the frame of the ``Spettroscopia laser e ottica
quantistica'' project of the Physics Deptartment of the University of Florence and
partially supported by Ente Cassa di Risparmio di Firenze and MIUR, under the FIRB
contract RBNE01KZ94. M.D. acknowledges the support of Marie Curie RTN.

\end{document}